\documentclass[reprint,
 superscriptaddress,
 bibnotes,
 amsmath,amssymb, aps,
 prl,
floatfix,
]{revtex4-1}

\usepackage{graphicx}
\usepackage[caption=false]{subfig}
\usepackage{dcolumn}
\usepackage{multirow}
\usepackage{tabularx}
\usepackage{bm}
\usepackage{hyperref}
\usepackage{amsmath}
\usepackage{amssymb}
\usepackage{comment}
\usepackage{color}
\usepackage{enumerate}
\usepackage{booktabs}
\usepackage{siunitx}

\newcommand{\Hz}{\text{~Hz}}

\def\apj{\ref@jnl{ApJ}}                 

\newcolumntype{L}[1]{>{\hsize=#1\hsize\raggedright\arraybackslash}X}%
\newcolumntype{R}[1]{>{\hsize=#1\hsize\raggedleft\arraybackslash}X}%
\newcolumntype{C}[1]{>{\hsize=#1\hsize\centering\arraybackslash}X}%

\makeatletter
\newcommand{\thickhline}{%
    \noalign {\ifnum 0=`}\fi \hrule height 2pt
    \futurelet \reserved@a \@xhline
}
\newcolumntype{"}{@{\hskip\tabcolsep\vrule width 2pt\hskip\tabcolsep}}
\makeatother

\newcommand{\Eq}[1]{Eq.~(\ref{#1})}

\newcommand{\Fig}[1]{Fig.~\ref{#1}}

\newcommand{\Ref}[1]{Ref.~\cite{#1}}

\begin{document}

\title{
Sound of Dark Matter: Searching for Light Scalars with Resonant-Mass Detectors
}

\author{Asimina Arvanitaki}
\email{aarvanitaki@perimeterinstitute.ca}
\affiliation{Perimeter Institute for Theoretical Physics, Waterloo, Ontario N2L 2Y5, Canada}
\author{Savas Dimopoulos} 
\email{savas@stanford.edu}
\affiliation{Stanford Institute for Theoretical Physics, Stanford University, Stanford, California 94305, USA}
\author{Ken Van Tilburg} 
\email{kenvt@stanford.edu}
\affiliation{Stanford Institute for Theoretical Physics, Stanford University, Stanford, California 94305, USA}

\date{\today}

\begin{abstract}
The fine-structure constant and the electron mass in string theory are determined by the values of scalar fields called moduli. If the dark matter takes on the form of such a light modulus, it oscillates with a frequency equal to its mass and an amplitude determined by the local dark-matter density. This translates into an oscillation of the size of a solid that can be observed by resonant-mass antennas. 
Existing and planned experiments, combined with a dedicated resonant-mass detector proposed in this Letter, can probe dark-matter moduli with frequencies between 1~kHz and 1~GHz, with much better sensitivity than searches for fifth forces. 
\end{abstract}
\maketitle


\textit{Introduction.---}
In string theory, the values of the fundamental parameters, such as the fine-structure constant or the electron Yukawa coupling, are functions of scalar fields called moduli. In a typical vacuum, there are several moduli that describe geometric properties of the extra dimensions of space, such as their size. The masses of the moduli are model dependent. Several moduli often remain massless as long as supersymmetry is unbroken.
 
For a supersymmetry-breaking scale near a TeV, the moduli can acquire a mass as large as $0.1~\text{meV}$ or a frequency of  $20~\text{GHz}$. 
These rough estimates are often corrected by small coefficients, such as loop, logarithmic, and large-volume factors, that make the moduli masses significantly lighter~\cite{Dimopoulos:1996kp,ArkaniHamed:1999dz, Burgess:2010sy, Cicoli:2011yy}.
Moduli associated with small numbers, such as the electron Yukawa coupling, are also naturally much lighter. One may even speculate that the ultrasmall cosmological constant is associated with an ultralight dilaton whose mass is of order the  Hubble scale~\cite{Damour:1994zq}. 

In the absence of a general theoretical mass range for moduli, we will only concern ourselves with experimental constraints. These scalars are an excellent dark matter (DM) candidate when produced through the misalignment mechanism. In order for a scalar to be a good DM candidate that gravitationally clumps at galactic scales, it has to be heavier than $10^{-22}~\text{eV}$~\cite{Hu:2000ke}. For scalar DM to be well characterized as a scalar field, instead of individual particles, its mass must be lighter than about $0.1~\text{eV}$.
Such a DM candidate---denoted by the field $\phi$---can cause fundamental constants to oscillate in time \cite{Arvanitaki:2014faa}.
We consider couplings to the electron $e$ and the electromagnetic field strength $F_{\mu \nu}$:
\begin{align}
\mathcal{L_\text{int}} \supset - \sqrt{4\pi G_N} \phi \left[ d_{m_e} m_e \bar e e - \frac{d_e}{4}F_{\mu\nu}F^{\mu \nu} \right], \label{eq:lagr}
\end{align}
where $G_N$ is Newton's constant ($\hbar = c = k_B = 1$ throughout). We can identify $\phi$ with an electron Yukawa (electric charge) modulus if $d_{m_e}~\neq~0$ ($d_e~\neq~0$). If $\phi$ constitutes the local DM energy density $\rho_\text{DM}$, it can be approximated by
\begin{align}
\phi(t,\mathbf{x}) \simeq \frac{\sqrt{2\rho_\text{DM}}}{m_\phi} \cos\left[m_\phi (t - \mathbf{v} \cdot \mathbf{x} + \dots)\right]\label{eq:phi},
\end{align}
where $|\mathbf{v}|$ is the relative velocity of the DM with respect to Earth, roughly equal to the virial velocity $v_\text{vir}$ in our Galaxy. 
The field oscillation occurs at an angular frequency equal to the DM mass, $m_{\phi}$, and exhibits high fractional temporal and spatial coherence of $v_\text{vir}^{-2} \sim 10^{6}$ and $v_\text{vir}^{-1}\sim 10^3$, respectively, due to a low velocity dispersion of the DM~\cite{calcaxion}. In such a background, the electron mass $m_e$ and the fine-structure constant $\alpha$ can fluctuate along with $\phi$ according to
\begin{align}
m_e(t,\mathbf{x}) &= m_{e,0} \left[ 1 + d_{m_e} \sqrt{4\pi G_N}\ \phi(t,\mathbf{x}) \right],\\
\alpha(t,\mathbf{x}) &= \alpha_0 \left[ 1 + d_e \sqrt{4\pi G_N}\ \phi(t,\mathbf{x}) \right].
\end{align}
The size of any atom is of order $1/\alpha m_e$, and will oscillate if $m_e$ or $\alpha$ fluctuate.
For a single atom, this is a tiny effect. However, it is enhanced when the atoms are stacked, as in a solid. 

In this Letter, we show how this amplification by the number of atoms, in combination with resonant effects, can be exploited to search for scalar DM with existing technology already used to search for gravitational wave (GW) radiation. In what follows, we explain how the signal arises, describe the reach of existing experiments, and discuss future directions, including a new experimental proposal. Finally, we compare with other constraints on scalar DM and find an improvement in sensitivity by several orders of magnitude.

\begin{figure*}[t]
\includegraphics[width = 0.99 \textwidth]{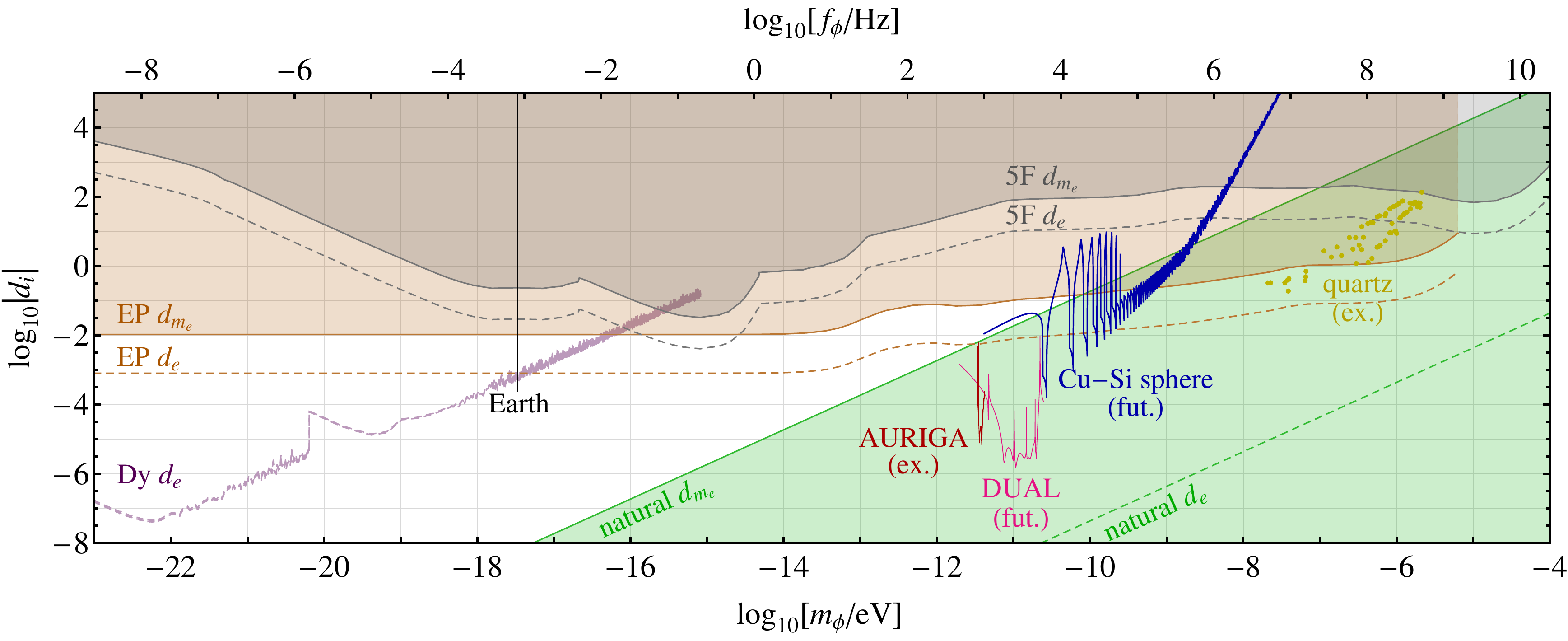}
\caption{Scalar field parameter space, with mass $m_\phi$ and corresponding DM oscillation frequency $f_\phi = m_\phi/2\pi$ on the bottom and top horizontal axes, and couplings of both an electron mass modulus ($d_i = d_{m_e}$) and electromagnetic gauge modulus ($d_i = d_e$) on the vertical axis.
Natural parameter space for a 10 TeV cutoff is depicted in green, while the other regions and dashed curves represent 95\% CL limits from fifth-force tests (``5F'', gray), equivalence-principle tests (``EP'', orange), atomic spectroscopy in dysprosium (``Dy'', purple), and low-frequency terrestrial seismology (``Earth'', black). The blue curve shows the projected $\text{SNR}=1$ reach of a proposed resonant-mass detector---a copper-silicon (Cu-Si) sphere 30 cm in radius---after $1.6~\text{y}$ of integration time, while the red curve shows the reach for the current AURIGA detector with $8~\text{y}$ of recasted data. Rough estimates of the 1-y reach of a proposed DUAL detector (pink) and several harmonics of two piezoelectric quartz resonators (gold points) are also shown. 
}\label{fig:coupling}
\end{figure*}

\textit{Experimental signature.---}
The effect of a DM modulus on a low-loss massive antenna can be captured by considering the response of a harmonic oscillator. An otherwise free mass $M$ on a physical, dissipative spring of equilibrium length $L$, resonant frequency $\omega$ and a quality factor $Q$ obeys
\begin{align}
M \left[\ddot x + \frac{\omega}{Q} \dot x + \omega^2 \left(x-L\right)\right] = F_{\text{th}} + F_{\text{ext}}. \label{eq:ODE1}
\end{align}
with thermal Brownian noise forces $F_{\text{th}}$ in the spring and external noise forces $F_{\text{ext}}$. In the presence of a modulus, the equilibrium size of the spring is oscillating in time $L \simeq {L}_0 \cos(m_\phi t)$. Once we define the ``displacement distance" $D \equiv x - L$, the influence of the modulus is revealed as a new force,
\begin{align}
M \left[\ddot{D}+ \frac{\omega}{Q} \dot{D} + \omega^2 D \right] \simeq - M \ddot L +  F_{\text{th}} + F_{\text{ext}}, \label{eq:ODE2}
\end{align}
up to $\mathcal{O}(1/Q)$-suppressed force terms. 
The modulus-induced force is analogous to the tidal force caused by a GW~\cite{Misner:1974qy}, except that the modulus induces a monopole strain instead of a quadrupole strain pattern. This intuition can be extended to continuous acoustic systems by describing the modulus as a scalar GW with an effective isotropic Riemann curvature tensor
\begin{align}
\mathcal{R}^\text{eff}_{i0j0} = \delta_{ij} \ddot h, \label{eq:riemann}
\end{align}
where the effective strain $h \equiv - \delta \alpha/ \alpha - \delta m_e/ m_e = -\left(d_{m_e} + d_e \right) \sqrt{4\pi G_N} \phi$ inherits the coherent properties of the DM field oscillation as described below \Eq{eq:phi}.
The response of a resonant-mass detector to modulus DM may thus be extracted from well-known strategies for detecting monochromatic gravitational-wave radiation.

A resonant-mass detector is acoustically equivalent to a combination of independent harmonic oscillators, since the displacement from equilibrium in an elastic solid can be decomposed into normal modes as $\mathbf{D}\left(\mathbf{x},t \right) = \sum_{n} D_{n}(t) \, \mathbf{u}_{n}(\mathbf{x})$.  In a spherical geometry, we can take $\mathbf{u}_n(\mathbf{x}) = \hat{r} u_n(r)$, since only spherically symmetric ($l=0$) modes are excited by a scalar strain. For a sphere of radius $R$ with uniform density $\rho$ and longitudinal (transverse) sound speed $c_l$ ($c_t$), these mode functions can be found in \Ref{landau1986theory}. They have resonant angular frequencies $\omega_n = c_l k_n$ with $k_n \simeq n \pi / R$. We choose a normalization such that $u_n(R) = 1$, and define the effective mode mass $M_n$ through $\int_\mathcal{V} d^3x \, \rho \, \mathbf{u}_n \cdot \mathbf{u}_{n'} = \delta_{nn'} M_n$. With these conventions, $D_n$ is the absolute displacement of the surface from the equilibrium radius $R_\text{eq}$, and satisfies \Eq{eq:ODE2} with an effective modulus force
\begin{align}
F_{\text{mod},n} \equiv  - \mathcal{R}_{i0j0} \int_\mathcal{V} d^3x \, \rho u_n^i x^j = - \ddot h M_\circ R J_n \label{eq:pseudoforce}
\end{align}
with $M_\circ$ the mass of the sphere and a coupling factor $J_n \equiv 3 R^{-4} \int_0^R dr \, r^3 u_n$ that decouples for the higher harmonics as $J_n \sim n^{-2}$. Modulus DM can be detected if the force in \Eq{eq:pseudoforce} exceeds the noise forces $F_\text{th}$ and $F_\text{ext}$.

\textit{Existing resonant-mass detectors.---}
The response of resonant-mass detectors to gravitational waves was first described by Weber~\cite{Weber:1960zz,Weber:1967}.
Resonant-mass GW detectors have made great strides in sensitivity since the first ``Weber bars" (see \Ref{aguiar1} for a historical review), so far culminating in a network of third-generation experiments~\cite{igec1,igec2} consisting of cryogenic, ton-scale, cylindrical antennas operating at around 900 Hz. Despite quality factors in excess of a million, these detectors achieve a sizable fractional bandwidth of $\mathcal{O}(10\%)$ by amplifying the surface displacement of the main antenna with a series of smaller mechanical and electrical resonators tuned to the frequency of the lowest longitudinal harmonic of the cylinder. The AURIGA Collaboration has achieved the widest bandwidth, operating at a noise level $S_{hh}^{1/2} \lesssim 10^{-20} \text{~Hz}^{-1/2}$ for $850 \text{~Hz} \lesssim f \lesssim 960 \text{~Hz}$ for an optimally polarized gravitational strain $h_{ij}$~\cite{auriga1}. Recasting as a projected reach for modulus couplings at unit signal-to-noise ratio ($\text{SNR} = 1$) with 8 years of data on tape yields the red curve in \Fig{fig:coupling}.

Astrophysical objects provide for naturally occurring resonant-mass antennas.
The fundamental breathing mode of Earth, which has a 20.46-min period and $Q\approx7500$, was studied with a dedicated seismometer in Ref.~\cite{JGR:JGR4118} over a 7-month period. The observed vertical acceleration noise spectrum of $7.6 \times 10^{-8} \text{~m}\text{~s}^{-2}\text{~Hz}^{-1/2}$ corresponds to a spherical strain sensitivity $S_{hh}^{1/2} \approx 8.4\times 10^{-14} \text{ Hz}^{-1/2}$ and a constraint $|h| \lesssim 4.5 \times 10^{-17}$. Interpreted as a 95\% CL limit on modulus DM couplings, this yields $|d_e + d_{m_e}| \lesssim 2.5 \times 10^{-4}$ in a $1.3 \times 10^{-4}$ fractional bandwidth around $f_\phi \approx 8.1 \times 10^{-4} \text{~Hz}$ (black line in \Fig{fig:coupling}).  
Modern examinations of Earth's higher harmonics~\cite{coughlin1} and crust excitations~\cite{dyson1,jensen1,coughlin2} may also be interesting, but likely have worse strain sensitivity.
Lunar~\cite{coughlin3} as well as asteroseismic observations~\cite{Lopes:2014dba,siegel1,Lopes:2015pca} have shown more promise towards detecting (quadrupole) metric variations;
monopole strain excitations of these systems merit further investigation.

\begin{figure}[t]
\includegraphics[width = 0.7 \columnwidth]{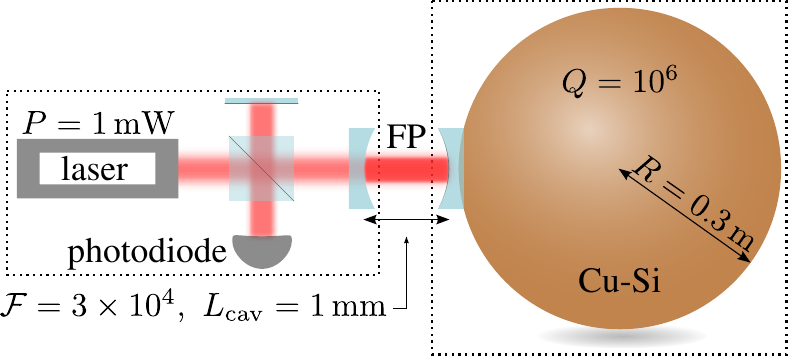}
\caption{Schematic of the proposed setup: a Cu-Si sphere whose surface displacement is monitored by a Fabry-P\'{e}rot (``FP'') interferometer. Elements encircled by the dotted lines are independently suspended and isolated from vibrations.
}\label{fig:setup}
\end{figure}

\begin{figure}[t]
\includegraphics[width = 0.99 \columnwidth]{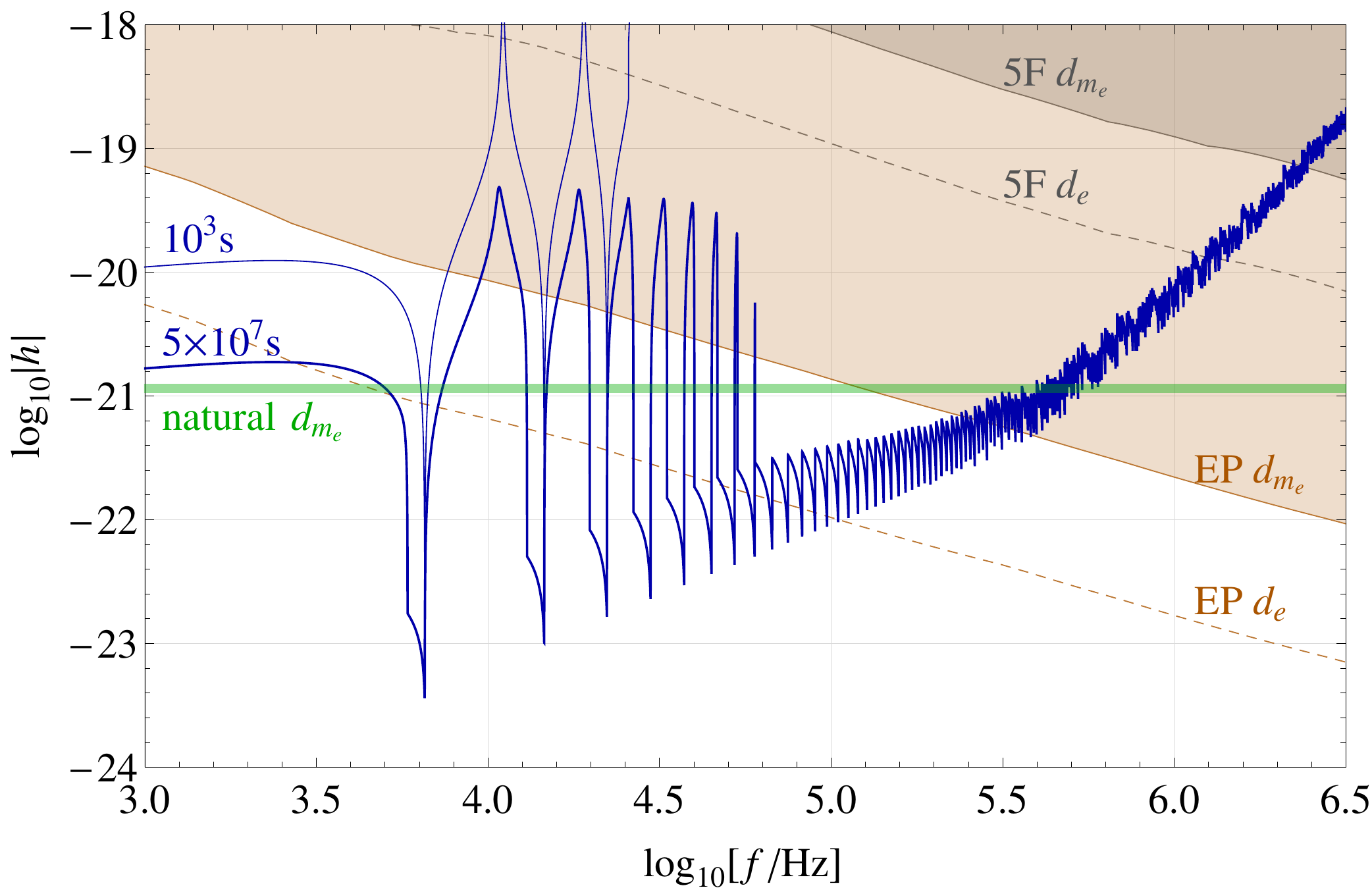}
\caption{Strain reach $|h|$ at $\text{SNR}=1$ as a function of frequency $f$ after an integration time $t_\text{int} = 5 \times 10^7 \text{~s}$ (thick blue curve), consisting of a 5\% fractional frequency scan by varying the temperature between 4~K and 100~K in increments, each $t_\text{shot} = 10^3 \text{~s}$ long. Reach for one ``shot'' at 4~K is shown by the thin blue curve (up to the third harmonic, for clarity). Equivalence-principle and fifth-force exclusions are shown in orange and gray, while strains below the green line are natural for an electron Yukawa modulus with a 10 TeV cutoff. 
}\label{fig:strain}
\end{figure}

\textit{Future resonant-mass detectors.---}
Given the unknown modulus mass, there is a clear need for a wideband detector, or a narrow-band one with scanning ability. Proposals in the former category, such as ``xylophone'' arrays \cite{xylo1,xylo2} and ``DUAL'' detectors \cite{dual1, dual2, dual3}, have been proposed for GW searches in the 1--10 kHz band, and would have excellent reach for moduli as well.
The wider-band DUAL proposal of \Ref{dual3} with a capacitive read-out may reach sensitivity to quadrupole strains at the level of $10^{-22}~\text{Hz}^{-1/2}$. The same noise spectrum for monopole strains would yield a reach as represented by the pink curve in \Fig{fig:coupling} after $1~\text{y}$ of data. (This is a useful proxy, because although a modulus would excite higher-frequency modes in that setup, we expect a comparable sensitivity with only minor read-out changes.) 

In this Letter, we propose a scanning experiment because of its simplicity and feasibility with current technology. The basic experimental concept, illustrated in \Fig{fig:setup}, is that of a freely suspended sphere, whose acoustic modes can be frequency-shifted by dialing the temperature and detected via an interferometric read-out of the sphere's surface displacement.

By exploiting the temperature dependence of elastic properties, adjustment of each mode's angular frequency $\omega_n = c_l k_n$ becomes possible. We propose a spherical antenna of radius $R = 0.3\text{~m}$ made out of a material such as C65500 copper-silicon alloy (high-silicon bronze ``A"), for which $c_l$ varies by about 5\% below 100 K while maintaining a high quality factor ($Q\sim 10^6$) \cite{duffy1,NBScopper}. Copper-based alloys are already used in spherical resonant-mass GW detectors \cite{minigrail1,schenberg1} for their high density $\rho \approx 8 \times 10^3 \text{~kg}/\text{m}^3$, sound speed $c_l \approx 4 \times 10^3 \text{~m}/\text{s}$, and thermal conductivity.

Brownian noise forces are broadband. By the fluctuation-dissipation theorem \cite{kubo}, their single-sided noise spectral density is $S^\text{th}_{FF} \simeq {4T M_n \omega_n}/{Q_n}$ around each mode, for a temperature $T \gg \omega_n$.
Using \Eq{eq:pseudoforce}, this translates to a near-resonance strain spectrum $S^\text{th}_{hh} \simeq  \frac{4 T R M_n}{M_\circ^2 Q_n c_l^3} \frac{1}{k_n^3 R^3 J_n^2 }$ or an amplitude of $1.5 \times 10^{-21} \Hz^{-1/2} (M_n/M_\circ k_n^3 R^3 J_n^2)^{1/2}$ at $4~\text{K}$, with the latter factor scaling as $\sim\hspace{-0.2em}n^{1/2}$.

A Fabry-P\'{e}rot interferometer, schematically drawn in \Fig{fig:setup}, can measure the sphere's total surface displacement $x = D + \delta R_\text{eq}$ through changes in the cavity length $L_\text{cav} = 1~\text{mm}$, which cause laser light 
to fluctuate in intensity on a photodiode. 
The shot-noise-limited displacement spectral density is $S^\text{ds}_{xx}(\omega) = S^\text{ds}_{xx,0}\left[1+(\omega/\Omega_\text{cav})^2 \right]$, where the cavity bandwidth is $\Omega_\text{cav} \equiv \pi/2 \mathcal{F} L_\text{cav}$ and the cavity finesse is taken to be $\mathcal{F}\approx 3 \times10^4$. 
We assume a baseline sensitivity of $S^\text{ds}_{xx,0} \approx 10^{-38} \text{~m}^2 \text{~Hz}^{-1} \sim \lambda/\mathcal{F}^2 P$, achievable with a laser power $P \sim 1 \text{~mW}$ and optical wavelength $\lambda$~\cite{hadjar}. 
Converting to a strain spectral density 
yields a near-resonance strain sensitivity of $S^\text{ds}_{hh}(\omega_n)^{1/2} \approx 10^{-25} \text{ Hz}^{-1/2} (M_n/M_\circ J_n)$ for $\omega_n \ll \Omega_\text{cav}$, scaling as $\propto n^2$ (above the cavity bandwidth, like $n^3$). Thermal noise sources in the interferometer can be kept subdominant above 1 kHz with fused silica mirror substrates, silica/tantala coatings, and a laser beam width of $1~\text{mm}$~\cite{Black:2004pf,flaminio1}. Vibration isolation schemes with $-200~\text{dB}$ attenuation exist for similar geometries~\cite{schenbergvib}, and should be able to reduce typical seismic noise spectra $\lesssim 10^{-10}~\text{m} ~\text{Hz}^{-1/2}$~\cite{araya1}
to negligible levels above $100~\text{Hz}$.
Monopole modes may be discriminated from multipole and other spurious modes via calibration hammer techniques~\cite{minigrail1} or multiple cavities~\cite{Bonaldi:2003ah}.

The reach of our proposal is shown in Figs.~\ref{fig:coupling}~\&~\ref{fig:strain}.
The resonant frequencies are adjusted in fractional increments of $10^{-6}$ (the fractional signal bandwidth) by varying $T$ in $2~\text{mK}$ steps. If each shot lasts a time $t_\text{shot} = 10^3~\text{s}$, 
a strain $h_\text{shot}(\omega,T_i) \simeq S_{hh}(\omega,T_i)^{1/2} t_\text{shot}^{-1/4} (2\pi /m_\phi v_\text{vir}^2)^{-1/4}$ can be detected at $\text{SNR}=1$, depicted by the thin blue curve in \Fig{fig:strain} for a temperature $T_i = 4~\text{K}$.
After an integration time $t_\text{int}=5 \times 10^7 ~\text{s}$, a set of $5 \times 10^4$ shots has a strain reach $h_\text{int}(\omega) \simeq \left[\sum_i h_\text{shot}(\omega,T_i)^{-4}\right]^{1/4}$ less than $10^{-20}$ in a $5\%$ band around each $l=0$ harmonic up to $n \sim 100$. 
Better sensitivity may eventually be attained at ultracryogenic temperatures. Since this precludes scanning of the resonant frequencies, traditional capacitive transducer read-outs as in Refs.~\cite{minigrail1,schenberg1} or more advanced interferometer schemes \cite{Marin:2003se} would have to be employed. 

Yet higher frequencies may be explored by micromechanical resonators, for which the unfavorable scaling of Brownian noise with size may be mitigated by using clever geometries and extremely low-loss materials. 
The proposal of \Ref{tobar1} to detect high-frequency GWs in curved-plate quartz crystals is also sensitive to modulus DM. An isotropic strain excites longitudinal acoustic modes, which, due to the piezoelectric nature of the quartz crystal, may be picked up by an electronic circuit. On-resonance strain sensitivities down to $10^{-22} ~\text{Hz}^{-1/2}$ are expected for up to a hundred modes per crystal~\cite{tobar1}. (Methods to increase the bandwidth are under development~\cite{tobar2}.) The modulus coupling reach of harmonics in two $20~\text{mK}$ sensors from \Ref{tobar1} is illustrated by the gold points in \Fig{fig:coupling}.

\textit{Non-acoustic experiments.---}
The phenomenology of light scalars with modulus couplings as in Eq.~\ref{eq:lagr} extends beyond the acoustic signature described above. 

Through scalar exchange, two macroscopic bodies with mass $M_1$ and $M_2$ experience a Yukawa force with a range set by $m_{\phi}^{-1}$. Its strength relative to gravity is $\alpha_\text{mod}^{(1,2)} \equiv  (d_1 Q^{}_{1})(d_2 Q^{}_{2}) $
with $d_i Q_{i} \equiv d_{m_e} Q_{m_e} + d_e Q_{e} $. We follow the notation of \Ref{Damour:2010rp}, in which $Q_{m_e}$ ($Q_e$) is the fractional amount of electron-mass (electromagnetic) energy relative to the rest-mass energy $M$ of the object. 

Searches for fifth forces at a length scale $L$ are sensitive to moduli with mass $m_\phi \sim 1/L$. A number of experiments have set constraints on composition-independent $|\alpha_\text{mod}| \lesssim 10^{-2.5}$ for $10^{-21} ~\text{eV} \lesssim m_\phi \lesssim 10^{-4.3}~\text{eV}$~\cite{Adelberger:2003zx}.
Rescaling by typical modulus charges $Q_{m_e} \sim 1/4000$ and $Q_e \sim 1/500$ yields approximate constraints on $|d_{m_e}|$ and $|d_e|$ shown in gray in \Fig{fig:coupling}.

The modulus force also violates the equivalence principle: two test masses $M_1$ and $M_2$ experience a different acceleration in the presence of a third one $M_3$ even though $M_1 = M_2$, provided that $\alpha^{(1,3)}_\text{mod} \neq \alpha^{(2,3)}_\text{mod}$. The E\"ot-Wash experiment \cite{Schlamminger:2007ht} has measured the fractional differential acceleration of beryllium and titanium in the Earth's gravitational field to be $(a_\text{Be} - a_\text{Ti})/a \approx (0.3 \pm 1.8)\times 10^{-13}$. Using  $Q^{\oplus}_i (Q^\text{Be}_i - Q^\text{Ti}_i) = \lbrace -2.42 \times 10^{-9}, -3.00 \times 10^{-6} \rbrace$ for $i = \lbrace m_e, e \rbrace$, one arrives at the 95\% CL upper limits $|d_{m_e}| \lesssim 1.05 \times 10^{-2}$ and $|d_e| \lesssim 2.98 \times 10^{-4}$ for $m_\phi \ll 1/R_\oplus$, shown in orange in \Fig{fig:coupling}. 
Bounds for $m_\phi \gtrsim 1/R_\oplus$ are estimated by rescaling the limit according to Ref.~\cite{Wagner:2012ui}. Lunar laser ranging sets less stringent constraints~\cite{Damour:2010rp,Williams:2004qba}.

Besides mediating EP-violating forces, a modulus field sourced by a massive body could also slightly alter fundamental constants around it. When sourced by the Sun, the modulus appears as an annual modulation of the fine-structure constant or the electron mass with a known phase, and with amplitude proportional to the Sun's modulus charges $Q_i^\odot$ and the $\pm1.65 \times 10^{-10}$ annual variation of the gravitational potential of the Sun on the Earth's orbit. The absence of such modulation in spectroscopy data of two different dysprosium isotopes~\cite{Leefer:2013waa} constrains $|d_e|$ to be less than $2.1 \times 10^{-2}$ for $m_\phi \ll (\text{AU})^{-1}$. Atomic clock pairs have the capability to greatly improve upon this technique, and extend it to other couplings.

Assuming that the field $\phi$ comprises part or all of the DM density, the modulus can be probed by spectroscopic searches for time-varying fundamental constants~\cite{Arvanitaki:2014faa}. Recently, Fourier analysis of transition energies in two dysprosium isotopes has set the tightest constraints on $|d_e|$ for $m_\phi \lesssim 3 \times 10^{-18} \text{~eV}$~\cite{VanTilburg:2015oza}, as indicated by the purple curve in \Fig{fig:coupling}. In the background field of \Eq{eq:phi}, a mass $M$ also experiences a (gradient-suppressed) force that may be observable in differential accelerometers such as free-mass GW detectors~\cite{Arvanitaki:2014faa}.

Black hole superradiance~\cite{Arvanitaki:2010sy} excludes scalars with $6 \times 10^{-13}~\text{eV} \lesssim m_\phi \lesssim 2 \times 10^{-11}~\text{eV} $ regardless of abundance and the couplings in  \Eq{eq:lagr}, unless $\phi$ has sufficiently strong self-interactions~\cite{Arvanitaki:2014wva}. Stellar cooling bounds are not competitive with force tests for the masses under consideration~\cite{Raffelt:1996wa}. A detailed summary of astrophysical constraints can be found in \Ref{Arvanitaki:2014faa}.

\textit{Discussion.---}
The mature technology of resonant-mass detectors provides a new way to probe scalar DM that couples to the electron mass and electric charge, with a sensitivity beyond that of EP and fifth-force tests, over a wide range of frequencies.

Resonant-mass detectors are better suited for scalar DM searches than free-mass interferometers such as LIGO, which have much reduced sensitivity to scalar GWs~\cite{Stadnik:2014tta} because of laser phase noise. By using equal-length interferometer arms, this noise can be canceled while leaving a quadrupole GW signal unaffected. However, this strategy also suppresses the scalar DM signal up to small gradient effects~\cite{Arvanitaki:2014faa}.

A scalar DM candidate frequently appears in string theory as a modulus, the dilaton or axions, even though for the latter the expected modulus couplings are far below our sensitivity levels~\cite{Arvanitaki:2014faa}.
Designing setups with wideband sensitivity is crucial given the unknown masses of these DM candidates. Theoretically, large couplings imply large radiative corrections to the mass of scalar particles, and thus bias towards a model-independent minimum mass: $m_\phi^2 \gtrsim \frac{1}{(4\pi)^3} d_{m_e}^2 y_e^2 G_N \Lambda^4 + \frac{1}{4\pi} d_e^2 G_N \Lambda^4$~\cite{Dimopoulos:1996kp}.
This naturalness criterion is satisfied inside the green bands of \Fig{fig:coupling} for a hard cutoff of $\Lambda \approx 10~\text{TeV}$.

The technology behind resonant-mass detectors has steadily improved over the past five decades, and offers a unique opportunity to search for a well-motivated DM candidate over a wide range of masses and couplings in the immediate future.

\acknowledgments{
We thank Andrew Geraci for advice regarding interferometric read-out schemes, and Massimo Cerdonio, Maxim Goryachev, Peter Graham, Leo Hollberg, Jeremy Mardon, Peter Michelson, Michael Tobar, and Robert Wagoner for fruitful discussions. Research at
Perimeter Institute is supported by the Government of Canada through Industry Canada and by the Province of Ontario through the Ministry of Economic Development \& Innovation. This work was partially supported by the National Science Foundation under grant no.~PHYS-1316699.
}


\bibliographystyle{apsrev4-1-etal}
\bibliography{SoundOfDM}

\end{document}